\documentclass[preprint,showpacs,superscriptaddress]{revtex4}
\usepackage{graphicx}

\newcommand{\bpm}{$B^\pm(V_G)$}
\newcommand{\dgt}{$dG_T/dV_G$}
\newcommand{\gtb}{$G_T(B)$}
\newcommand{\dtwof}{$\partial^2f/\partial V_G^2$}
\newcommand{\figsch} {Fig.~\ref{Schematic}}
\newcommand{\figvgb} {Fig.~\ref{VGB}}
\newcommand{\figgam} {Fig.~\ref{Gamma}}
\newcommand{\figdmn} {Fig.~\ref{Diamonds}}

\begin{document}
\author{H. Steinberg} \affiliation{Dept. of Condensed Matter Physics,
Weizmann Institute of Science, Rehovot 76100, Israel}
\author{O.~M. Auslaender*} \affiliation{Dept. of Condensed Matter Physics,
Weizmann Institute of Science, Rehovot 76100, Israel}
\author{A. Yacoby} \affiliation{Dept. of Condensed Matter Physics, Weizmann Institute
of Science, Rehovot 76100, Israel}
\author{J. Qian}\affiliation{Lyman Laboratory of Physics, Harvard University, Cambridge, MA 02138, USA}
\author{G.~A. Fiete}\affiliation{Lyman Laboratory of Physics, Harvard University, Cambridge, MA 02138, USA}\affiliation{Kavli Institute for Theoretical Physics, University of California, Santa Barbara, CA 93106}
\author{Y. Tserkovnyak}\affiliation{Lyman Laboratory of Physics, Harvard University, Cambridge, MA 02138, USA}
\author{B.~I. Halperin}\affiliation{Lyman Laboratory of Physics, Harvard University, Cambridge, MA 02138, USA}
\author{K.~W. Baldwin} \affiliation{Bell Labs, Lucent Technologies, 700 Mountain Avenue,
Murray Hill, NJ 07974, USA.}
\author{L.~N. Pfeiffer} \affiliation{Bell Labs, Lucent Technologies, 700 Mountain Avenue,
Murray Hill, NJ 07974, USA.}
\author{K.~W. West} \affiliation{Bell Labs, Lucent Technologies, 700 Mountain Avenue,
Murray Hill, NJ 07974, USA.}

\title{Localization Transition in a Ballistic Quantum Wire}

\begin{abstract}
The many-body wave-function of an interacting one-dimensional
electron system is probed, focusing on the low-density, strong
interaction regime. The properties of the wave-function are
determined using tunneling between two long, clean, parallel
quantum wires in a GaAs/AlGaAs heterostructure, allowing for
gate-controlled electron density. As electron density is lowered
to a critical value the many-body state abruptly changes from an
extended state with a well-defined momentum to a localized state
with a wide range of momentum components. The signature of the
localized states appears as discrete tunneling features at
resonant gate-voltages, corresponding to the depletion of single
electrons and showing Coulomb-blockade behavior. Typically $5-10$
such features appear, where the one-electron state has a
single-lobed momentum distribution, and the few-electron states
have double-lobed distributions with peaks at $\pm k_F$. A
theoretical model suggests that for a small number of particles
($N<6$), the observed state is a mixture of ground and thermally
excited spin states.
\end{abstract}

\date{\today}


\maketitle

%begin text

Coulomb interactions in many-body quantum systems can lead to the creation
of exotic phases of matter. A prime example is a Luttinger liquid, which describes
 a system of interacting electrons confined to one spatial dimension
\cite{Haldane81}. At high electron densities the electron kinetic
energy dominates over the Coulomb energy and the transport
properties of the system resemble those of non-interacting
electrons. In this weakly interacting limit, conductance is
quantized even in the presence of moderate disorder
\cite{Tarucha95,Yacoby96}. Reducing the electron density
suppresses the kinetic energy more rapidly than the Coulomb
energy, leading to the strongly interacting limit, where charge
correlations resemble those in a Wigner crystal, an ordered
lattice of electrons with periodicity $n^{-1}$ ($n$, the mean
electron density). In this limit, one expects the weakest amount
of disorder to pin the crystal, thereby suppressing conductance at
low temperatures \cite{Glazman92}. Here we investigate the
suppression of conductance and the changes in the electron
wavefunction as $n$ is reduced, using momentum-resolved tunneling.
We find a remarkably sharp localization transition at low
densities.

Momentum resolved tunneling between two quantum wires has been
shown to be an effective experimental tool in the study of
interacting one-dimensional (1D) systems. This method uses
tunneling across an extended junction between two closely situated
parallel clean quantum wires \cite{Ophir02}. An electron tunneling
across the junction transfers momentum $eBd\equiv\hbar q_B$ ($B$
is the magnetic field perpendicular to the plane of the wires, $d$
is the distance between them, as in \figsch). When the source
drain bias voltage $V_{SD}$ applied across the junction is low,
the probability for an electron to tunnel between the wires can be
measured through the low-bias tunneling conductance \gtb\
\cite{OphirHadar05}. At low temperatures $G_T(B)\propto |M(k)|^2$,
assuming the lower wire is uniform and weakly interacting, where
$M(k)$ is the tunneling matrix element. $M(k)$ is given by
\begin{equation}
\label{Mk}
M(k)=\int_{-\infty}^{\infty}dxe^{ikx}\Psi(x),
\end{equation}
where $\Psi(x)$ is a ``quasi-wavefunction" for the upper wire,
defined by: $\Psi(x)=\left<N-1|\psi(x)|N\right>$. Here
$k=q_B-k_F^L$, $k_F^U, k_F^L$ are the Fermi-wavenumbers in the
Upper, Lower wires (UW, LW), $|N\rangle$ is the ground state for
$N$ particles in the UW, and $\psi(x)$ is an operator that removes
an electron from point $x$ in the UW \cite{Fiete05}. At finite
temperatures, several states contribute, and the conductance is
proportional to a weighted average of their corresponding
$|M(k)|^2$. In the absence of interactions, $\Psi(x)$ would be the
wave function of the $N$th electron, which is a plane-wave for an
infinite system. In this case we intuitively expect
$|M(k)|^2\propto \delta(k + k_F^U) + \delta(k - k_F^U)$. \gtb\ is
therefore finite only when $|B|=B^{\pm}$, with
\begin{equation}
\label{BPlusMinus} B^{\pm}=\frac{\hbar}{ed}\left|k_F^U\pm
k_F^L\right|
\end{equation}
allowing tunneling only between the Fermi-points of the two wires.

The precise line shape of \gtb\ encodes details of the microscopic
properties of the many-body states involved in the tunneling
process: A realistic, finite size junction introduces effects
manifest as fringes accompanying the $\delta$-function peaks of
$M(k)$ \cite{Tserkovnyak02}. Electron-electron interactions reduce
the norm of $\Psi(x)$, but many of its qualitative features remain
unchanged for clean, long wires at low temperatures: \gtb\ should
have peaks at the same values of $B$ as in the non-interacting
case. We use \gtb\ to probe the real-space structure of $\Psi(x)$
in the presence of confinement and electron interactions. We
report on measurements of \gtb\ as a function of $n$, following it
down to the strongly interacting, low-density regime.

The experimental setup is schematically drawn in \figsch. It is
realized by two parallel wires at the edges of two quantum wells
fabricated in a GaAs/AlGaAs heterostructure by cleaved edge
overgrowth (CEO) \cite{Pfeiffer93}. Only the top quantum well
($0.5\mu m$ deep) is populated by a two-dimensional electron gas
(2DEG) which serves as a contact to the UW through ohmic contacts
$O_{1,2,3}$. The experiment is set up using $2 \mu m$-wide
tungsten gates on the top surface $G_{1,2,3}$: $G_3$ is set to a
negative voltage where the UW is depleted and LW remains
continuous, so that only the tunneling current is measured between
source $O_1$ and drain $O_3$. $G_1$ is set to deplete both wires,
so that the tunneling junction at the drain is much longer than
the one at the source, ensuring that the LW is kept at drain
potential. The density of a segment in the wire is varied by
applying voltage $V_G$ to $G_2$. $O_2$ is set to drain voltage to
allow simultaneous reading of 2-terminal UW conductance and
tunneling conductance. The density-dependent measurement consists
of setting $B$ and measuring the tunneling current $I_T=G_TV_{SD}$
as a function of $V_G$. The tunneling current includes
contributions which are not $V_G$ dependent. To single out the
density-dependent contribution, the differential tunneling
conductance \dgt\ is measured using a lock-in. Typically $dV_G$ is
a few $mV$ with a $4Hz$ frequency. The measurements are performed
in a $^3He$ refrigerator at temperatures down to $0.25K$.

Initially a cooled sample has one populated sub-band in the UW
(henceforth ``single-mode wire"). The potential landscape along
the UW is marked by $U_S(x)$ in \figsch, where the depletion of
the wire requires a relatively small negative gate voltage
($-0.9V$). Illuminating the sample with infrared light increases
the electron density in the wire thus allowing further population
of higher sub-bands (``multi-mode wire"). Depletion of the wire
after illumination requires a larger ($-3.5V$) gate voltage. The
finite slope in $U_M(x)$ results in a shorter effective length of
the low-density region.
\begin{figure}

\includegraphics[width=3.375 in,angle=0,clip=]{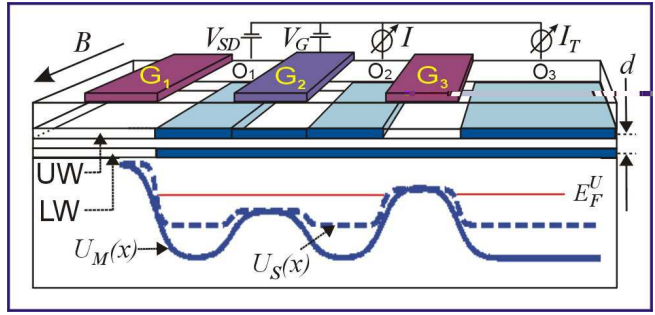}
\caption{\label{Schematic} A schematic of the measurement setup
with cleave plane front, perpendicular to $B$. Depicted: $2 \mu
m$-wide top gates ($G_1$, $G_2$ and $G_3$), $20 nm$-thick upper
wire at the edge of the 2DEG, $30 nm$-thick lower wire and $6 nm$
insulating AlGaAs barrier. $U_S(x) (U_M(x))$: Schematic of UW
gate-induced potentials for single-mode (multi-mode) wires,
$E_F^U$ is the Fermi energy of the upper wire. Ohmic contact $O_1$
serves as source, $O_{2,3}$ as drains. Density is controlled by
gate voltage $V_G$ on gate $G_2$. Two-terminal current is marked
by $I$, tunneling current by $I_T$.}
\end{figure}

\begin{figure}[ptb]
\includegraphics[width=3.375 in,angle=0,clip=]{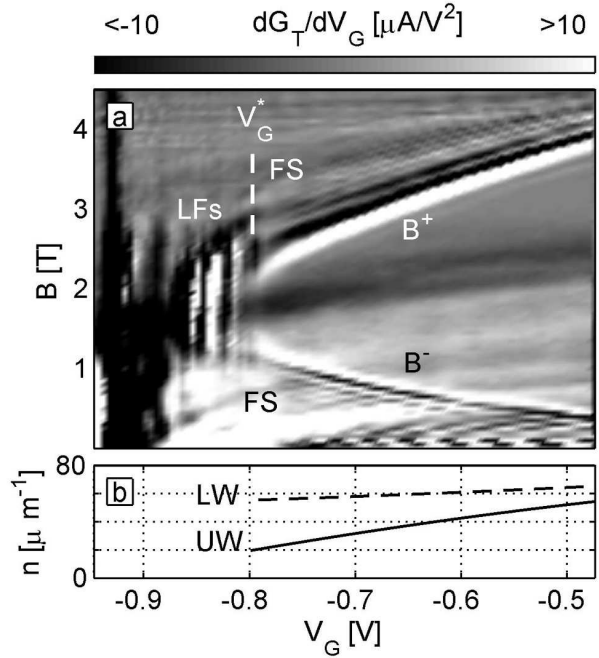}
\caption{\label{VGB} (a) Plot of \dgt\ versus $V_G$ and $B$ for a
single-mode wire. Applied bias $V_{SD}=100\mu V$ selected to avoid
the zero-bias anomaly \cite{Tserkovnyak03}, but is small enough to
consider tunneling between the Fermi-points of both wires. The
upper and lower curves are momentum conserving tunneling features
$B^\pm(V_G)$. Each curve is accompanied by Finite Size features
(FS). At low densities localization features (LFs) appear instead
of these curves. $V_G^*$ marks the localization transition. (b) UW
and LW densities extracted using Eq.~\ref{BPlusMinus}.}
\end{figure}

\figvgb\ (a) shows the measurement of \dgt\ for a single-mode
wire. The figure is dominated by a set of curves marked \bpm. The
upper curve $B^+(V_G)$ corresponds to $+$ sign in
Eq.~\ref{BPlusMinus} and is the measured differential conductance
which results from tunneling between counter-propagating states.
The lower curve corresponds to the $-$ sign in
Eq.~\ref{BPlusMinus} which in turn results from tunneling between
co-propagating states. Each curve is accompanied by Finite Size
fringes (FS) which are a consequence of the finite length of the
low-density region. As the top-gate voltage grows more negative
the density under the gate is reduced and the curves converge.
Panel (b) shows the densities in both wires as extracted using
Eq.~\ref{BPlusMinus}.
%We note that for $V_G >
%-0.9$ the 2DEG is not yet depleted and may appear in the tunneling
%signature. Initially $n_{2D} \approx 10^{11} cm^{-2}$ which may
%fit the faint signals appearing for $B < 2T$.

An abrupt change in \dgt\ is evident in \figvgb\ at a critical
gate-voltage $V_G^*=-0.80V$, which corresponds to a critical UW
density $n^*=20 \mu m^{-1}$. At $n^*$ the high density features
\bpm\ disappear, giving way to a different set of features
labelled Localization Features (LFs). This is a fundamental
change: When $n>n^*$ the wire is uniform and momentum is a good
quantum number. \dgt\ is therefore appreciable only in a narrow
range of $\hbar/edL_U$ around $B^\pm$, $L_U$ being the length of
the density-tuned region. When $n<n^*$ The momentum range spanned
by the LFs is typically very broad and lies roughly between the
extrapolations of the \bpm\ curves. Further attesting to the
importance of this transition, we note that simultaneous reading
of $I$ and $I_T$ (\figsch) shows that this change is concurrent
with the suppression of the 2-terminal conductance
\cite{OphirHadar05}.

The broad momentum spectrum exhibited by the LFs may have several
possible interpretations. For example, $\Psi(x)$ may be localized
on a length scale not much larger than the inter-particle
separation. Alternatively, broad LFs may result either from
non-uniform density which gives way to a broad distribution of
local Fermi wavevectors, or from contributions of states above the
ground state, with excitations that soak up part of the
transferred momentum.

The LFs are studied more closely in \figgam\, where
high-resolution scans of single-mode and multi-mode wire LFs are
presented (panels (a) and (b) respectively). The LFs appear as
vertical streaks in $B$. They are narrow in the $V_G$ direction,
separated by strips with vanishing signal. Such discrete behavior
is a result of charge quantization typical of Coulomb blockade
(CB) of electrons in a finite box. The number of the LFs for
several cases is summarized in Table~\ref{LTable}.

The CB behavior, along with the observation that the UW states are
localized, implies the existence of a localized region separated
from the rest of the UW by tunnel-barriers. This localized region
forms a quantum-dot with three leads: Two leads are the UW
segments which are not density controlled, and one is the LW. It
is not clear, however, why in this regime we typically don't
detect resonances in the UW conductance. As expected from CB
physics, when a finite source-drain bias is applied (\figdmn), the
LFs split to form the well known diamonds \cite{Devoret92}. The
asymmetry of the diamonds indicates different capacitance to each
lead.

\begin{figure}[ptb]
\includegraphics[width=3.375 in,angle=0,clip=]{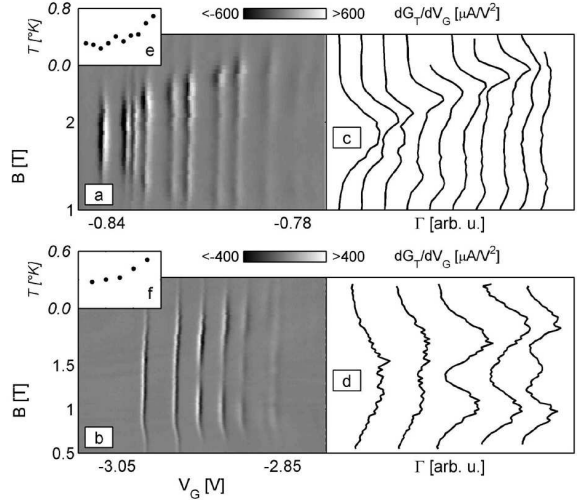}
\caption{\label{Gamma} (a) High resolution measurement of \dgt\ of
Localization Features for a single-mode wire, $V_{SD}=50\mu V$,
$dV_G=300\mu V$. (b) Same as (a), taken for the 2nd sub-band in a
multi-mode wire. (c) $\Gamma(B)$ of panel (a) LFs.  (d)
$\Gamma(B)$ of panel (b). (e) $T(N)$ of panel (a), $N$ is the
number of the added electron for each LF. (f) $T(N)$ of panel (b).
$\Gamma$ and $T$ are defined in the text.}
\end{figure}

We define the length of the low-density region above and below
$n^*$ as $L_{ex}$ and $L_{loc}$ respectively. Both are extracted
from the data: In the extended regime, $n>n^*$, finite size
fringes accompany the $B^\pm(V_G)$ curves for values of
$B<B^-(V_G)$ and $B>B^+(V_G)$, (marked by ``FS" in \figvgb). The
location of the FS, outside the $B^{\pm}$ curves, implies that the
electron density here has a minimum in the center of the UW, with
a length $L_{ex} \approx 2 \pi / \Delta q_B $,  $\Delta q_B$ being
the spacing of the fringes in the $B$ direction
\cite{Tserkovnyak02}. In the localized regime, $n<n^*$,
$L_{loc}=N/n^*$, where $N$ is the number of electrons confined to
the low-density region (number of LFs). Table~\ref{LTable}
summarizes the results for the LFs presented in this paper and in
\cite{OphirHadar05}. We see that $L_{ex}/L_{loc} \approx 2$. This
suggests that the entire low-density region participates in the
localization.

\begin{table}
\begin{tabular}{|c|c|c|c|c|c|}
  \hline
  Sub-band:&$N$ &$n^* [\mu m^{-1}]$ &$L_{loc} [\mu m]$ &$L_{ex} [\mu m]$ &$V_G^*$ \\
  \hline
  Single-Mode & 12 & 20 &  0.6 & 1.0 & -0.9 \\
  Multi-Mode 2nd &  5 & 22 & 0.23 &     & -2.6 \\
  Multi-Mode 1st &  6 & 25 &  0.24 & 0.5 & -3.6 \\
  \hline
\end{tabular}
\caption{\label{LTable} Summary of $L_{loc}$, $L_{ex}$, $N$ and
 $V_G^*$ for a single-mode wire and modes $1,2$ of
a multi-mode wire}
\end{table}

\figdmn\ contains a line-scan showing the cross section of the
peaks and a fit to a second derivative of the Fermi-function,
\dtwof\ using two adjustable parameters: The width of the peak,
scaling linearly with temperature, and its height, which is
proportional to the tunneling rate, $\Gamma$. Such fits were
performed for each value of $B$ in \figgam. The extracted values
of $\Gamma$ are presented in \figgam (c) and (d). The
Coulomb-blockaded states can now be characterized by
$\Gamma(B)\propto |M(k)|^2$, which is a direct measure of their
momentum distribution (Using Eqs.~1 and 2). The single-mode
measurement presented in panel (c) shows a generic case where
$\Gamma(B)$ has a single peak for the first LF $(N=1)$ and two
broad peaks for $N\geq 2$, with a separation $\Delta B$ that
increases as N is increased. There is only little variation in
this behavior upon different cool-downs, single vs. multi-mode
conditions and when a different gate is used. Rarely one can find
momentum distribution that can be interpreted using single
particle theory. \figgam (d) shows the LF's of subband 2 in a
multi mode wire. One can clearly see that $N=1,2$ are
approximately single-lobed, $N=3,4$ double-lobed, and $N=5$ is
triple lobed. This behavior is characteristic of the the three
first spin degenerate wave function in a one dimensional well
\cite{Fiete05}.

Insets (e) and (f) of \figgam\ show the medians of the temperature
fit for each of the LFs in (a) and (b) respectively. The temperature
is originally obtained in units of $V_G$ and is translated to $K$
using the capacitance ratio measured in scans such as \figdmn. The
deduced temperature appears to increase with the addition of
electrons, almost by a factor of $2$. The apparent broadening of the
feature can not be explained by level broadening since the
line-shape does not resemble a Lorentzian. Furthermore, scans such
as \figdmn\ rule out the possibility that the features broaden due
to changing of lead-dot capacitance. We currently do not have a
straightforward explanation for this observation.

\begin{figure}[ptb]
\includegraphics[width=3.375 in,angle=0,clip=]{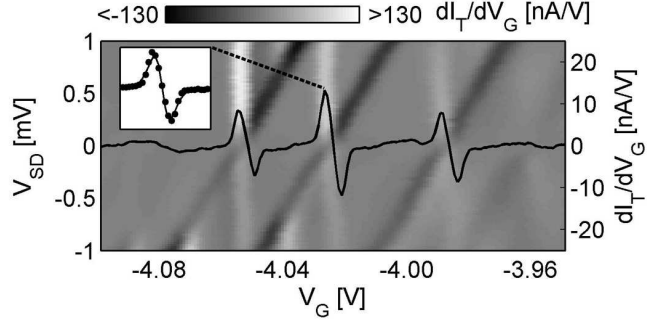}
\caption{\label{Diamonds} $dI_T/dV_G$ as a function of gate
voltage $V_G$ and source-drain bias $V_{SD}$, taken for the
Localization Features of sub-band 1 in a multi-mode wire, at
$B=2T$ . The tilted diamond features are due to resonant-tunneling
through the localized states. A line-scan of $dI_T/dV_G$  at a
bias of $20 \mu V$ is superimposed on the plot. Inset: 2nd
derivative of a Fermi-function (line) and data (dots) of the 2nd
LF.}
\end{figure}

In an attempt to model the data, we have calculated the expected
tunneling form factor $|M(k)|^2$ in the localized regime: For
$N=1$, $\Psi(x)$ has no nodes. This agrees with the experimental
results in \figgam\ (b) and (d). For $N=2$, however, where the
spin-singlet ground state would reproduce a similar signature, the
observed $\Gamma(B)$ deviates from the single-lobe structure. We
therefore estimate the energy difference between the lowest
triplet state and the singlet ground state for the two electron
system and find it to be extremely small, less than the Zeeman
energy in a field of $2T$. Mixing of the singlet-triplet states at
finite temperatures reproduces the observed structure: A pair of
maxima at finite $|k|$ with a non-zero local minimum at $k=0$
\cite{Fiete05}.

The energy scale for spin-excitations grows rapidly with
increasing density, and at $T=0.25K$ we expect spin excitations to
be frozen out for $N \ge 5$. The estimate is sensitive to the form
and size of the cutoff in the interaction potential. Calculations
assuming sharp confinement lead to a form factor $|M(k)|^2$ which
becomes sharply peaked near $k = \pm k_F^U$. The broad peaks
observed experimentally and the substantial weight near $k = 0$
could be a result of softness or asymmetry in the confining
potential, some residual disorder, or thermally excited
spin-excitations (possibly due to the exchange energy being
smaller than our estimate) \cite{Cheianov04, Fiete04, Fiete05}.

In conclusion, we have studied the momentum-structure of the
many-body wave-function of an interacting quantum-wire. We find
that below the critical density $n^*$, where the 2-terminal
conductance is suppressed, the electrons are bound to a localized
region and display Coulomb-blockade charging physics. Momentum
spectroscopy of the localized few-electron states reveals an
evolution of momentum structure from single-peak to double-peak.

We would like to thank R. de-Picciotto for his contribution. Y.
Oreg and T. Giamarchi assisted in helpful discussions. This work
was supported in part by the US-Israel BSF, the European
Commission RTN Network Contract No. HPRN-CT-2000-00125 and NSF
grants DMR-02-33773 and PHY-01-17795. YT is supported by the
Harvard Society of Fellows. GAF is supported by NSF grant
PHY-99-07949 and by the Packard Foundation. OMA acknowledges
partial support from a grant from the Israeli Ministry of Science.
HS is supported by a grant from the Israeli Ministry of Science.

%end text
${}*$ Present address: Gaballe Laboratory for Advanced Materials,
Stanford University, Stanford, CA 94305, USA.

%\bibliography{hadar}

\bibliographystyle{apsrev}

\end{document}